\documentclass[12pt,epsfig]{article}
\usepackage{epsfig}
\usepackage{amssymb}

\begin{document}

%%%%%%Define some new commands and macros

\newcommand{\beq}{\begin{equation}}
\newcommand{\eeq}{\end{equation}}
\newcommand{\bea}{\begin{eqnarray}}
\newcommand{\eea}{\end{eqnarray}}
\newcommand{\beqn}{\begin{eqnarray}}
\newcommand{\eeqn}{\end{eqnarray}}
\newcommand{\beas}{\begin{eqnarray*}}
\newcommand{\eeas}{\end{eqnarray*}}
\newcommand{\defi}{\stackrel{\rm def}{=}}
\newcommand{\non}{\nonumber}
\newcommand{\bquo}{\begin{quote}}
\newcommand{\enqu}{\end{quote}}
\newcommand{\p}{\partial}

%%%%%%%%%%%%%%%%%%%%%%%%%%%%%%%%%% definitions

\def\de{\partial}
\def\Tr{ \hbox{\rm Tr}}
\def\const{\hbox {\rm const.}}
\def\o{\over}
\def\im{\hbox{\rm Im}}
\def\re{\hbox{\rm Re}}
\def\bra{\langle}\def\ket{\rangle}
\def\Arg{\hbox {\rm Arg}}
\def\Re{\hbox {\rm Re}}
\def\Im{\hbox {\rm Im}}
\def\diag{\hbox{\rm diag}}

\def\stroke{\vrule height8pt width0.4pt depth-0.1pt}
\def\topfleck{\vrule height8pt width0.5pt depth-5.9pt}
\def\botfleck{\vrule height2pt width0.5pt depth0.1pt}
\def\Zmath{\vcenter{\hbox{\numbers\rlap{\rlap{Z}\kern
0.8pt\topfleck}\kern
2.2pt\rlap Z\kern 6pt\botfleck\kern 1pt}}}
\def\Qmath{\vcenter{\hbox{\upright\rlap{\rlap{Q}\kern
3.8pt\stroke}\phantom{Q}}}}
\def\Nmath{\vcenter{\hbox{\upright\rlap{I}\kern 1.7pt N}}}
\def\Cmath{\vcenter{\hbox{\upright\rlap{\rlap{C}\kern
3.8pt\stroke}\phantom{C}}}}
\def\Rmath{\vcenter{\hbox{\upright\rlap{I}\kern 1.7pt R}}}
\def\Z{\ifmmode\Zmath\else$\Zmath$\fi}
\def\Q{\ifmmode\Qmath\else$\Qmath$\fi}
\def\N{\ifmmode\Nmath\else$\Nmath$\fi}
\def\C{\ifmmode\Cmath\else$\Cmath$\fi}
\def\R{\ifmmode\Rmath\else$\Rmath$\fi}

%%%%%%%%%%%%%%%%%%%%%%%%%%%%%%%%%%%%%%%%%%%%%%%%%%%%%%%%%%%%%%%%%%%%

\def\QATOPD#1#2#3#4{{#3 \atopwithdelims#1#2 #4}}
\def\stackunder#1#2{\mathrel{\mathop{#2}\limits_{#1}}}
\def\stackreb#1#2{\mathrel{\mathop{#2}\limits_{#1}}}
\def\Tr{{\rm Tr}}
\def\res{{\rm res}}
\def\Bf#1{\mbox{\boldmath $#1$}}
\def\balpha{{\Bf\alpha}}
\def\bbeta{{\Bf\beta}}
\def\bgamma{{\Bf\gamma}}
\def\bnu{{\Bf\nu}}
\def\bmu{{\Bf\mu}}
\def\bphi{{\Bf\phi}}
\def\bPhi{{\Bf\Phi}}
\def\bomega{{\Bf\omega}}
\def\blambda{{\Bf\lambda}}
\def\brho{{\Bf\rho}}
\def\bsigma{{\bfit\sigma}}
\def\bxi{{\Bf\xi}}
\def\bbeta{{\Bf\eta}}
\def\d{\partial}
\def\der#1#2{\frac{\d{#1}}{\d{#2}}}
\def\Im{{\rm Im}}
\def\Re{{\rm Re}}
\def\rank{{\rm rank}}
\def\diag{{\rm diag}}
\def\2{{1\over 2}}
\def\ntwo{${\cal N}=2\;$}
\def\4N{${\cal N}=4$}
\def\none{${\cal N}=1\;$}
\def\x{\stackrel{\otimes}{,}}
\def\beq{\begin{equation}}
\def\eeq{\end{equation}}
\def\beqn{\begin{eqnarray}}
\def\eeqn{\end{eqnarray}}
\def\ba{\beq\new\begin{array}{c}}
\def\ea{\end{array}\eeq}
\def\be{\ba}
\def\ee{\ea}
\def\stackreb#1#2{\mathrel{\mathop{#2}\limits_{#1}}}

\def\baselinestretch{1.0}

\begin{titlepage}

\begin{flushright}
FTPI-MINN-05/49\\
UMN-TH-2421/05\\
ITEP-TH-106/05\\
%December 13, 2005
\end{flushright}

\vspace{1cm}

\begin{center}

{\Large \bf The Higgs and Coulomb/Confining  Phases in
``Twisted-Mass" Deformed \boldmath{$CP(N-1)$} Model}
\end{center}

\vspace{0.5cm}

\begin{center}
{{\bf A.~Gorsky}$^{\,a,b}$, {\bf M.~Shifman}$^{\,b}$ and
{\bf A.~Yung}$^{\,a,b,c}$}
\end {center}
\begin{center}

$^a${\it Institute of   Theoretical and Experimental Physics,
Moscow  117250, Russia}\\
$^b${\it  William I. Fine Theoretical Physics Institute,
University of Minnesota,
Minneapolis, MN 55455, USA}\\
$^c${\it Petersburg Nuclear Physics Institute, Gatchina, St.
 Petersburg
188300, Russia}\\

\end{center}

\vspace{3mm}

\begin{abstract}

We consider non-supersymmetric two-dimensional $CP(N-1)$ mo\-del
deformed by
a term presenting the  bosonic part of the twisted mass deformation
of  \ntwo supersymmetric version of the model. Our deformation has a
special form preserving a $Z_N$ symmetry at the Lagrangian level.
In the large mass limit the
model is weakly coupled. Its dynamics is described by the Higgs
phase,
with $Z_N$ spontaneously broken. At small
masses it is in the strong coupling Coulomb/confining phase.
The $Z_N$ symmetry is restored.
Two phases are separated by a phase transition. We find the phase
transition point in the large-$N$ limit. It lies at strong coupling.
As was expected, the phase transition is related
to broken versus unbroken $Z_N$ symmetry in these two respective
phases.
The vacuum energies for these phases are determined too.

\end{abstract}

\end{titlepage}
%%%%%%%%%%%%%%%%%%%%%%%%%%%%

\section{Introduction}
\label{one}

As well known, two-dimensional $CP(N-1)$ model is an excellent
theoretical laboratory for modeling,
in a simplified environment, a variety of interesting phenomena
typical of  non-Abelian gauge theories in four dimensions
\cite{W79,NSVZsigma}.
Recently, two-dimensional $CP(N-1)$ model was shown to emerge
\cite {GSY05} as a moduli theory on the world-sheet of non-Abelian
flux tubes
presenting solitons in certain four-dimensional Yang--Mills theories
at weak coupling
\cite{HT1,ABEKY,SYcm,HT2}.
The flux tube solutions in the bulk (``microscopic") Yang--Mills
theory
depend on an adjustable parameter of dimension of mass. When this
parameter is large
the flux tubes are in fact $Z_N$ strings; they evolve towards
non-Abelian strings
as the above mass parameter decreases and eventually vanishes.
Correspondingly, the world-sheet theory is not just the $CP(N-1)$
model;
rather it is the $CP(N-1)$ model mass-deformed in a special way that
preserves a $Z_N$
symmetry of the model. The mass term we  deal with coincides with a
special choice of the twisted mass
\cite{Alvarez} in  supersymmetric  $CP(N-1)$ model. We hasten to
emphasize
that we will focus exclusively on non-supersymmtric version, to which
the two-phase phenomenon we study is inherent. There is no such
phenomenon in supersymmetric version.

In the limit of vanishing mass deformation, the
$CP(N-1)$ model is known to be a strongly coupled asymptotically free
field theory
\cite{P75}. A dynamical scale $\Lambda$ is generated as a result of
dimensional transmutation.
However, at large $N$ it can be solved by virtue of $1/N$ expansion
\cite{W79}.
The solution found by Witten exhibits a composite photon,
coupled to $N$ quanta $n$, each with charge $1/\sqrt N$ with respect
to this photon. In two dimensions the corresponding potential is long-range.
It causes linear confinement, so that only $n^*n$ pairs show up in the
spectrum. This is the reason why we refer to this phase as
``Coulomb/confining." In the Coulomb/confining phase the vacuum 
is unique and the $Z_N$ symmetry is unbroken.

On the other hand, if the mass deformation parameter is $\gg
\Lambda$,
the model is at weak coupling, the field $n$ develops a vacuum
expectation value (VEV),
there are $N$ physically equivalent vacua, in each of which the
$Z_N$ symmetry is spontaneously broken. We will refer to this regime
as the
Higgs phase, although this name has a Pickwick sense. Usually
the Higgs mechanism implies that a gauge boson eats a would-be
Goldstone meson
thus acquiring a mass that screens long range interactions.
In our case, at $m\gg \Lambda$, there is no gauge boson to begin
with (see Sect.~\ref{four}).
However, the long-range interaction inherent to the
Coulomb/confining phase does not take place; that's why it is not 
unreasonable
to refer to the phase as the Higgs phase.

In Ref. \cite{GSY05}
it was argued that the twisted mass deformed $CP(N-1)$ model
undergoes a phase transition when the value of the mass parameter is
$\sim \Lambda$.
The argument was largely based on analysis of the flux tubes and
their evolution in the underlying four-dimensional theory. In this paper we will
show, basing our consideration  on two-dimensional model {\em per se}, in
the large $N$ limit,
that a phase transition between $Z_N$ broken and unbroken (i.e. the
Higgs and Coulomb/confining) phases does indeed occur at
$m^2=\Lambda^2.$
The change of regimes takes place in a narrow interval
$m^2-\Lambda^2=O(1/N)$
where the method we use is insufficient to resolve details of the
phase transition.
In particular, the task of finding a  conformal field theory emerging
at the critical point
$m^2=\Lambda^2$ remains open.

The issue of two phases and phase transitions in related models
was previously addressed by Ferrari
\cite{Ferrari,Ferrari2}. While the first paper \cite{Ferrari}
deals with $CP(N-1)$ models,
neither the methods used nor results have a
significant overlap with the results reported below. In \cite{Ferrari2}
Ferrari exploits $1/N$ expansion methods which are similar to
ours. The model to which Ref.~\cite{Ferrari2} is devoted is a mass-deformed
$O(N)$ model with a $Z_2$ symmetry at the Lagrangian level. The point of the 
phase transition separates the $Z_2$ broken phase at weak coupling from the
the $Z_2$ unbroken phase at strong coupling, which is in parallel with our 
result.
The phase transition in \cite{Ferrari2} is argued to be of the Ising-model 
type.
We do not expect this to be valid in the $Z_N$ case we deal with in the 
present paper.

The paper is organized as follows. In Sect.~\ref{two}
we introduce the twisted mass deformation of the non-supersymmetric
$CP(N-1)$ model that preserves $Z_N$ at the Lagrangian level. We also
review
some well-known facts regarding this model in the strong and weak
coupling regimes.
The weak coupling regime (large twisted masses) is especially simple
since here the theory can be treated perturbatively and exhibits a
Higgs-like behavior.
At strong coupling we are guided by Witten's large-$N$ solution.
In Sect.~\ref{three} large-$N$ methods are used to solve the model at
arbitrary $m$.
The critical point
separating $Z_N$ broken and unbroken phases
is determined and the vacuum energies are calculated for both phases.
Section \ref{four} presents a remark concerning
inadequacy of the  large-$N$ methods for determination  of
the nature of the critical behavior.

\section{Twisted-mass deformed \boldmath{$CP(N-1)$} model and its
phases}
\label{two}

In this section we describe a twisted mass deformation of the  
$CP(N-1)$ model
preserving $Z_N$. Then we discuss its two distinct phases in two
opposite limits, $m \ll\Lambda $ and $m\gg\Lambda$.

As was mentioned, the origin of the word ``twisted" lies in 
supersymmetry, more exactly,
extended \ntwo supersymmetry. Aspects of the supersymmetric version
were analyzed by Dorey \cite{Dorey},
who found an exact solution in the holomorphic sector.
We study the non-supersymmetric version,
obtained by discarding the fermion sector.

\subsection{The model}
\label{themodel}

As usual in two dimensions, the Lagrangian can be cast in many
different (but equivalent) forms. For our purposes the
most convenient formulation is in terms of the $n$
fields.\footnote{They are referred to as ``quarks" or solitons in
Ref.~\cite{W79}.}
To set our notation, let us first omit the twisted mass.
Then the $CP(N-1)$ model can be written as
\beq
S = \int d^2 x \left \{
\left(\partial_{\alpha} + i A_\alpha\right) n^*_{\ell}
\left(\partial_\alpha - i A_\alpha \right) n^{\ell}
+\lambda \left( n^*_{\ell} n^{\ell}-r\right)
\right \}\,,
\label{Tone}
\eeq
where $n^\ell$ is an $N$-component complex filed,
$\ell = 1,2,...,N$, subject to the constraint
 \beq
n_{\ell}^*\, n^{\ell} =r\,
\label{lambdaco}
\eeq
where $r$ is the inverse coupling constant of the model.
More exactly, the standard relation between $r$ and $g^2$ is
$$
r = {2}/{g^2}\,.
$$
The action (\ref{Tone}) and other similar expressions below are given
in the Euclidean space.
The constraint (\ref{lambdaco}) is implemented by the Lagrange
multiplier $\lambda$
in Eq.~(\ref{Tone}). The field $A_\alpha$ in the Lagrangian is
auxiliary too;
it enters with no derivatives and can be eliminated
by virtue of the equation of motion,
\beq
A_\alpha =-\frac{i}{2r}\, n^*_\ell \stackrel{\leftrightarrow}
{\partial_\alpha} n^\ell\,.
\label{Ttwo}
\eeq
Substituting Eq.~(\ref{Ttwo}) in the Lagrangian, we rewrite the
action in
the form
\beq
S = \int d^2 x  \left\{
\partial_{\alpha}  n^*_{\ell}\,
\partial_\alpha  n^{\ell} + \frac1{r}
(n_\ell^*\partial_\alpha n^\ell)^2
+\lambda \left( n^*_{\ell} n^{\ell}-r \right)
\right\}\,.
\label{Tthree}
\eeq

\vspace{2mm}

The model (\ref{Tthree}) is a generalization of the O(3) sigma model.
The latter is formulated as
\beq
S = \frac{r}{4}\,\int d^2 x \, \partial_{\alpha}\vec S \,
\partial_{\alpha}\vec S
\label{Tfour}
\eeq
where $\vec S$ is a three-component vector
subject to the constraint $\vec S\,^2 =1$. At $N=2$ the $CP(N-1)$ model
(i.e. $CP(1)$) reduces to O(3) through the substitution
\beq
S^a = \frac{1}{r}\left(  n^*\,\tau^a\, n\right) ,\qquad a=1,2,3\,,
\label{Tfive}
\eeq
where $\tau^a$ are the Pauli matrices.
The constraint $\vec S\,^2 =1$ follows from Eq.~(\ref{lambdaco}),
while
\beq \partial_{\alpha}\vec S \, \partial_{\alpha}\vec S
\leftrightarrow
\frac4r\,\left\{\partial_{\alpha}  n^*_{\ell}\,
\partial_\alpha  n^{\ell} +
\frac1r(n_\ell^*\partial_\alpha n^\ell)^2\right\}\,.
\label{Tsix}
\eeq

\vspace{2mm}

The coupling constant $r$ is asymptotically free \cite{P75}, and
defines
the dynamical scale of the theory $\Lambda$ through
\beq
\Lambda^2 = M_{\rm uv}^2 \exp\left(-\frac{4\pi r_0}{N}\right)\,,
\label{Tseven}
\eeq
where $M_{\rm uv}$ is the ultraviolet cut-off and $r_0$
is the bare coupling.
The combination $N/r$ is nothing but
the 't Hooft constant which does not scale with $N$.
As a result, $\Lambda$ scales as $N^0$  at large $N$.
One can also introduce the $\theta$ term, if one  so desires,
\beq
S_\theta =\frac{i\theta}{2\pi }\, \int d^2 x \,
\varepsilon_{\alpha\gamma}\,
\partial^\alpha A^\gamma
= \frac{\theta}{2\pi r}\,  \int d^2 x \, \varepsilon_{\alpha\gamma}
\left(\partial^\alpha  n_{\ell}^*\,\, \partial^\gamma n^{\ell}
\right).
\label{Teight}
\eeq

\vspace{2mm}

Now let us add to the action (\ref{Tone}) or (\ref{Tthree}) a mass
term
of a special form,
\beq
S_m = \int d^2x \, \sum_\ell\left\{
n^*_\ell (\sigma^* - m^*_\ell)(\sigma - m_\ell)n^\ell
\right\}\,,
\label{twmass}
\eeq
where $\sigma$ is an auxiliary complex field (with no kinetic term),
and we choose
\beq
m_\ell = m \exp\left(\frac{2\pi i\, \ell}{N}\right)\,,\qquad \ell =
0,1,..., N-1
\,.
\label{m}
\eeq
The parameter $m$ in Eq.~(\ref{m}) can be assumed to be real
and positive. This is not the most general choice of the
twisted mass deformation. In general, a single condition is imposed,
$\sum_\ell m_\ell =0$, which destroys SU($N$)/U(1) preserving only
residual U(1)$^{N-1}$.
We want to maintain an additional $Z_N$ symmetry, however, which is
automatic
under (\ref{m}). The $Z_N$ symmetry of the action has historic roots
\cite{GSY05}, but what is important at present is that the
twisted mass deformed model with  $Z_N$ symmetry  is
interesting on its own. The $Z_N$ symmetry plays an important role
in identifying a phase transition between the  Higgs and
Coulomb/confining
phases of the theory.

Eliminating $\sigma$ by virtue of the equation of motion,
\beq
\sigma = \frac1r \sum_\ell\, m_\ell\, n^*_\ell \, n^\ell\,,
\label{Ttwelve}
\eeq
we get
\beq
S_m =  \int d^2 x\, \,  m^2\left\{ r -\frac1r \left|\sum_\ell
 e^{\frac{2\pi i \ell}{N}}
 n^*_\ell \, n^\ell
\right|^2
\right\}\,.
\label{Tthirteen}
\eeq
It is instructive to see what becomes of this mass term at $N=2$ (i.e. the O(3) sigma model).
Then, Eq.~(\ref{Tthirteen}) implies
\beq
S_m =\, r \int d^2 x
\, m^2 \,(1-S_3^2)\,,
\label{wb}
\eeq
which is obviously $Z_2$ symmetric.
If $m$ is large, $m\gg\Lambda$,
the theory has two vacua, at $S_3=1$ and $S_3=-1$.
In both vacua there are two elementary excitations,
$S_1$ and $S_2$, with masses $2m$.

In the general $N$ case the action at hand has the following $Z_N$
symmetry:
\beqn
&&\sigma \to  e^{i\frac{2\pi k}{N}}\sigma,\qquad
n_\ell \to  n_{\ell +k}\, \nonumber\\[2mm]
&& \mbox{for every fixed}\,\,\,\ell\,\,\,
\mbox{and}\,\,\,
k=1,2,...,N\,.
\label{cycle}
\eeqn

\subsection{The Higgs phase}
\label{thehiggs}

At large $m$, $m\gg\Lambda$, the renormalization group flow of the
coupling constant is frozen at the scale $m$.
Thus, the model at hand is at weak coupling and  the
quasiclassical analysis is applicable.
The potential  (\ref{twmass}) have $N$ degenerate vacua which are
labeled by the order parameter $\langle\sigma\rangle$,
the vacuum configuration being
\beq
\sigma = m_{\ell_0}\,,\qquad n^{\ell_0}=\sqrt{r}\,, \quad\mbox{and}
\quad n^\ell =0\,\,\,\mbox{if}\,\,\,
\ell\neq \ell_0\,.
\label{higgsvac}
\eeq
In each given vacuum
the $Z_N$ symmetry (\ref{cycle}) is spontaneously
broken.

There are $2(N-1)$ elementary excitations\,\footnote{Here we count
real degrees of
freedom. The action (\ref{Tone}) contains $N$ complex fields
$n^\ell$.
The phase of $n^{\ell_0}$ can be eliminated from the very beginning.
The condition $n_\ell^*n^\ell = r$ eliminates one more field.
} with physical masses
\beq
M_\ell = |m_\ell-m_{\ell_0}|\,,\qquad \ell\neq\ell_0\,.
\label{elmass}
\eeq
Besides, there are kinks (domain ``walls" which are particles in two
dimensions) interpolating between these vacua. Their masses scale as
\beq
M^{\rm kink}_{\ell} \sim r\,M_\ell \,.
\label{kinkmass}
\eeq
The kinks  are much  heavier than elementary
excitations at weak coupling. Note that they have nothing to do
with Witten's $n$ solitons \cite{W79} identified as solitons at
strong coupling. The point of phase transition separates these 
two classes of solitons.

\subsection{The Coulomb/confining phase}
\label{thecoulomb}

Now let us discuss the Coulomb/confining phase of the theory
occurring at small $m$.
As was mentioned, at $m=0$ the $CP(N-1)$ model was solved by Witten
in the
large-$N$ limit \cite{W79}. The model at small $m$ is very similar to
Witten's solution. (In fact, in the large-$N$ limit it is just the same.)
In Sect.~\ref{theccp} we present a generalization of
Witten's analysis which we will use to study the phase transition
between the $Z_N$ asymmetric and symmetric phases.
Here we just briefly summarize Witten's results for
the massless model.

If $m=0$, classically the  field $n^{\ell}$ can have arbitrary
direction; therefore, one might naively expect
spontaneous breaking of SU($N$) and
the occurrence of massless Goldstone modes. Well, this cannot happen
in two dimensions. Quantum effects restore the
full symmetry making the vacuum unique. Moreover, the condition
(\ref{lambdaco}) gets in effect relaxed. Due to strong coupling
we have more degrees of freedom than in the original Lagrangian,
namely all $N$ fields $n$ become dynamical and acquire
masses $ \Lambda$.

\begin{figure}
\epsfxsize=8cm
\centerline{\epsfbox{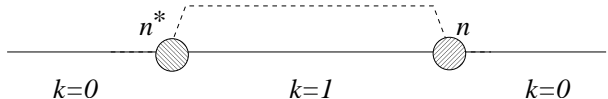}}
\caption{
Linear confinement of the $n$-$n^*$ pair.
The solid straight line represents the ground state.
The dashed line shows
the vacuum energy density (normalizing ${\cal E}_0$ to zero).}
\label{fig:conf}
\end{figure}

This is not the end of the story, however. In addition, one gets
another
composite degree of freedom.
The  U(1) gauge field $A_{\alpha}$ acquires a standard
kinetic term at  one-loop level,\footnote{By loops here we mean
perturbative expansion in $1/N$ perturbation theory.
} of the form
\beq
{N}\,\Lambda^{-2}\,\,  F_{\alpha\beta}  \,  F_{\alpha\beta}.
\label{gkinterm}
\eeq
Comparing Eq.~(\ref{gkinterm}) with (\ref{Tone})
we see that the charge of the $n$ fields with respect to this photon
is $1/\sqrt{ N}$.
The Coulomb potential between two charges in
two dimensions is linear in separation  between these charges.
The linear potential scales as
\beq
V(R) \sim \frac{\Lambda^2}{N}\, R
\label{nine}
\eeq
where $R$ is  separation.
The force is attractive for pairs
$\bar n$ and $n$, leading to the formation
of weakly coupled bound states (weak coupling is the
manifestation of the $1/N$ suppression of the confining potential).
Charged states are eliminated from the spectrum. This is the reason
why the
$n$ fields were called ``quarks" by Witten. The spectrum
of the theory consists of $\bar{n} n$-``mesons.'' The picture of
confinement of $n$'s is shown in Fig.~\ref{fig:conf}.

The validity of the above consideration rests on large $N$.
If $N$ is not large the solution \cite{W79}
ceases to be applicable. It remains  valid in the qualitative sense,
however. Indeed, at $N=2$ the model was solved exactly \cite{ZZ,Zam} (see also
\cite{Coleman}).
Zamolodchikovs found that the spectrum of the
O(3) model consists of a triplet of degenerate states
(with mass $\sim \Lambda$).
At $N=2$ the action (\ref{Tthree}) is built of doublets.
In this sense one can
say that Zamolodchikovs' solution exhibits confinement of
doublets. This is in qualitative accord with the large-$N$ solution
\cite{W79}.

Inside the $\bar n\,n$ mesons, we have  a constant electric field,
see Fig.~\ref{fig:conf}. Therefore the spatial interval between $\bar n$
and $n$ has a higher energy density than the domains outside the meson.

Modern understanding of the vacuum structure of the massless
$CP(N-1)$ model
\cite{nvacym} (see also \cite{nvacymp})
allows one to reinterpret  confining dynamics of the
$n$ fields in different terms \cite{MMY,GSY05}. Indeed, at large $N$,
along with the unique ground state,
the model has $\sim N$ quasi-stable local minima, quasi-vacua,
which become absolutely stable at $N=\infty$. The relative
splittings between the values of the energy density in the adjacent
minima
is of the order of $1/N$, while
the probability of the false vacuum decay is proportional to
$N^{-1}\exp (-N)$ \cite{nvacym,nvacymp}. The $n$
quanta ($n$ quarks-solitons) interpolate between
the adjacent minima.

The existence of a large family of quasi-vacua
can be inferred from the study of the $\theta$ evolution of the
theory.
 Consider the topological susceptibility, i.e. the correlation
function of two topological densities
\beq
\int d^2 x \, \langle Q(x),\,\, Q(0)\rangle\,,
\label{correlator}
\eeq
where
\beq
Q=\frac{i}{2\pi }\, \varepsilon_{\alpha\gamma}
\partial^\alpha A^\gamma
=\, \frac{1}{2\pi r}\, \varepsilon_{\alpha\gamma}
\left(\partial^\alpha
n_{\ell}^*\,\, \partial^\gamma n^{\ell}
\right).
\label{topdens}
\eeq
The correlation function (\ref{correlator}) is proportional to the
second derivative
of the vacuum energy with respect to the $\theta$ angle. From
(\ref{topdens})
it is not difficult to deduce that this correlation function  scales
as $1/N$ in the large $N$
limit. The vacuum energy by itself scales as $N$. Thus, we conclude
that, in fact, the
vacuum energy should be a function of $\theta/N$.

\begin{figure}
\epsfxsize=6cm
\centerline{\epsfbox{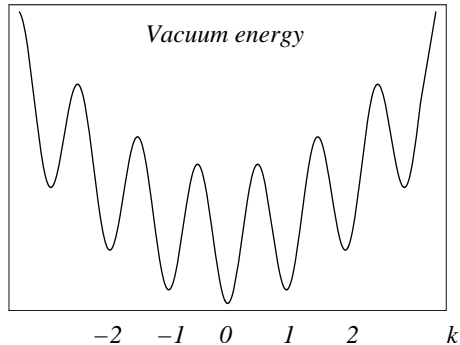}}
\caption{
The vacuum structure of
$CP(N-1)$ model at $\theta =0$.}
\label{odin}
\end{figure}

On the other hand, on general grounds,  the vacuum energy must be
a $2\pi$-periodic function of
$\theta$. These two requirements are seemingly self-contradictory.
A way out reconciling the above facts is as follows. Assume that we
have a family of quasi-vacua
 with energies
\beq
 E_k (\theta) \sim \, N\, \Lambda^2
\left\{1 + {\rm const}\left(\frac{2\pi k +\theta}{N}
\right)^2
\right\}
\,,\,\,\,\, k=0\ldots, N-1
\label{split}
\eeq
A schematic picture of these vacua
is given in Fig. ~\ref{odin}. All  these minima are entangled in the
$\theta$ evolution.
If we vary $\theta$ continuously from $0$ to $2\pi$ the
depths of the minima ``breathe." At $\theta =\pi$ two vacua become
degenerate, while for larger values of $\theta$ the former
global minimum
becomes local while the adjacent local minimum becomes global.
It is obvious that for the neibohring vacua which are not too far
from the global minimum
\beq
E_{k+1}-E_k \sim \frac{\Lambda^2}{N}\,.
\label{distance}
\eeq
This is also the confining force acting between $n$ and $\bar n$.

One could introduce  order parameters that would
distinguish between distinct vacua from the vacuum family.
An obvious choice is the expectation value of the topological charge.
The kinks $n^{\ell}$ interpolate, say, between the global minimum
and the first local one on the right-hand side. Then $\bar n$'s
interpolate between the
first local minimum and the global one. Note that the vacuum energy
splitting
is an effect suppressed by $1/N$.
At the same time, kinks  have masses which scale
as $N^0$,
\beq
M^{n\,\,{\rm kink}}_{\ell}\sim \Lambda\,.
\label{kinkmasscoulomb}
\eeq
The multiplicity of
such kinks is $N$ \cite{Acharya}, they form an $N$-plet of SU($N$).
This is in full accord with the fact that the large-$N$
solution of (\ref{Tone}) exhibits $N$ quanta of the complex field
$n$.

In  summary, the  $CP(N-1)$
model in the Coulomb/confining phase,
at small $m$, has a
vacuum family with a fine structure.
For each given $\theta$ (except $\theta =\pi, \,\, 3\pi$, etc.)
the true ground state is unique, but there is a large number
of ``almost" degenerate ground states.
The splitting is of the order of $\Lambda^2/N$. The $Z_N$ symmetry is
unbroken. The spectrum
of physically observable states consists of kink-anti-kink  mesons
which form  the   adjoint representation of SU($N$).

At large
$m$ the theory is in the Higgs phase; it has $N$ strictly degenerate
vacua; the $Z_N$ symmetry is broken. We have
$N-1$ elementary excitations $n^{\ell}$  with  masses
given by Eq.~(\ref{elmass}).

Thus we conclude that these two regimes
should be separated by a phase transition \cite{GSY05}. This phase
transition is associated with the $Z_N$ symmetry breaking: in the
Higgs phase the
$Z_N$ symmetry is spontaneously broken, while in the Coulomb phase it
is restored. For $N=2$ we deal with $Z_2$ which makes the situation
akin to the
Ising model.

\section{Solution at \boldmath{$m\neq 0$} in the large-$N$ limit}
\label{three}

In this section we will generalize Witten's analysis \cite{W79}
to include $m\neq 0$. The twisted mass deformed action is
\beq
S = \int d^2 x \left \{
|\nabla_{\alpha} n^{\ell}|^2
+\lambda \left( | n^{\ell}|^2-r_0\right)
+ \sum_\ell |(\sigma - m_\ell)n^\ell|^2
\right \}\,,
\label{model}
\eeq
were $\nabla_{\alpha}=\partial_{\alpha}-
iA_{\alpha}$ and $m_\ell$ is defined in Eq. (\ref{m}), and $r_0$
is the bare coupling constant.

\subsection{Effective theory}
\label{effective}

As soon as the action (\ref{model}) is quadratic in $n^{\ell}$
we can integrate over these fields and then minimize the resulting
effective
action with respect to other fields. Large-$N$ limit ensures that
corrections to the saddle point approximation are small. In fact,
this procedure boils down to calculating  one-loop graphs with fields
$n^{\ell}$ propagating inside loops.

In the Higgs phase the field $n^{\ell_0}$ develop a VEV.
One can always choose $\ell_0=0$ and denote $n^{\ell_0}\equiv n$. 
The field $n$, along with $\sigma$, are
our order parameters that distinguish between the
Coulomb/confining and the Higgs phases, see (\ref{higgsvac}).

Therefore, we do not want
to integrate over $n$ {\em a priori}. Instead,
we will stick to the following strategy:  we integrate over $N-1$
fields $n^{\ell}$ with $\ell \ne 0$.
The resulting effective action is to be considered as
a functional of $n$, $\lambda$ and
$\sigma$. To find the vacuum configuration, we will minimize the
effective action with
respect to $n$, $\lambda$ and
$\sigma$.

Integration over $n^{\ell}$
 with $\ell \ne 0$ produces the  determinant
\beq
\prod_{\ell=1}^{N-1}
\left[ {\rm det}\, \left(-\p_{\alpha}^2 +\lambda
+|\sigma-m_{\ell}|^2\right)
\right]^{-1},
\label{det}
\eeq
where we dropped the gauge field $A_\alpha$. In principle, as was
explained in
Sect.~\ref{thecoulomb},
quasi-vacua in the Coulomb/confining phase have non-vanishing
expectation
values of the operator (\ref{topdens}). However, we cannot see these
VEVs in
the leading order in $N$. Since the analysis we carry out applies to
the leading order in
the large-$N$ limit, we set $A_{\alpha}=0$.

Calculating (\ref{det}) we get the following contribution to the
effective action:
\beq
\frac1{4\pi}\sum_{\ell=1}^{N-1}\left(\lambda
+|\sigma-m_{\ell}|^2\right)
\left[\ln\, {\frac{M_{\rm uv}^2}{\lambda
+|\sigma-m_{\ell}|^2}}+1\right],
\label{detr}
\eeq
where we dropped a quadratically divergent contribution which does
not depend on
$\lambda$ and $\sigma$.

Equation (\ref{Tseven}) implies that the bare coupling constant
 $r_0$ in (\ref{model}) can be parameterized as
\beq
r_0=\frac{N}{4\pi}\, \ln\, {\frac{M_{\rm uv}^2}{\Lambda^2}}\,.
\eeq
Substituting this expression in (\ref{model}) and adding (\ref{detr})
we see that the term proportional to 
$\lambda \ln\, {M_{\rm uv}^2}$ is canceled out, and the effective action is
expressed in terms of the renormalized coupling constant,
\beq
r_{\rm ren}=\frac{1}{4\pi}\, \sum_{\ell=1}^{N-1}
\ln\, {\frac{\lambda +|\sigma-m_{\ell}|^2}{\Lambda^2}}\, ,
\label{coupling}
\eeq
where we neglect $O(1/N)$ contributions.
In addition to the coupling constant renormalization  we have to
carry out
renormalization of the   field $\sigma$
leading to a renormalization of its vacuum expectation value.
To this end we add the corresponding  counterterm to the bare  action
(\ref{model}), namely,
\beq
-\frac1{4\pi}\sum_{\ell=1}^{N-1}|\sigma -m_{\ell}|^2 \left(
\ln\, {\frac{M^2_{\rm uv}}{\Lambda^2}} -c\right),
\label{counterterm}
\eeq
where $c$ is a finite constant to be fixed below. This counterterm
ensures
 that the infinite term proportional to
$\sum_{\ell=1}^{N-1}|\sigma -m_{\ell}|^2\ln\, {M^2_{\rm uv}}$ in the
determinant (\ref{detr}) is canceled and, the renormalized VEV of
$\sigma$
is finite. We fix the coefficient $c$ below demanding
that   $\langle\sigma\rangle -m_{0} = 0$ in the Higgs phase, see 
(\ref{higgsvac}).

Assembling  all contributions together we get the effective action in
the form
\beqn
 S &=& \int d^2 x \Big\{
|\p_{\alpha} n|^2
+\left(\lambda +  |\sigma - m_0|^2\right)|n|^2
\nonumber\\[3mm]
&+&\frac1{4\pi}\, \sum_{\ell=1}^{N-1}\left(\lambda
+|\sigma-m_{\ell}|^2\right)
\left[1-\ln\, {\frac{\lambda +|\sigma-m_{\ell}|^2}{\Lambda^2}}\right]
\nonumber\\[3mm]
&+& \left.
\frac1{4\pi}\,
\sum_{\ell=1}^{N-1}|\sigma -m_{\ell}|^2\,c\right\}\,.
\label{effaction}
\eeqn
Now, minimizing this action with respect    $\lambda$, $n$ and
$\sigma$ we arrive at the following set of equations:
\beqn
&&
|n|^2=r_{\rm ren}\,,
\label{leq}
\\[3mm]
&& \left(\lambda +  |\sigma - m_0|^2\right)n=0\,,
\label{neq}
\\[3mm]
&&
-\frac1{4\pi}\,
\sum_{\ell=1}^{N-1}\left(\sigma-m_{\ell}\right)\,
\ln\, {\frac{\lambda +|\sigma-m_{\ell}|^2}{\Lambda^2}}
+(\sigma -m_0)\,
|n|^2 +\frac{N}{4\pi}\,c\,\sigma=0\, ,\nonumber\\
\label{seq}
\eeqn
where $r_{\rm ren}$ is given in Eq.~(\ref{coupling}), and we take
into account the fact that the sum $\sum_{\ell=1}^{N-1}m_{\ell}$
is relatively suppressed: instead of $O(N)$ it is $O(1)$ so that
we lose the factor of $N$.
Equations (\ref{leq}),
(\ref{neq}) and (\ref{seq}) represent our master set that
 determines the vacua of the theory.
Note that Eq. (\ref{leq}) is a renormalized version of the
bare condition (\ref{lambdaco}). In addition to the above equations
the vacuum configuration must satisfy two
 extra constraints,
\beq
r_{\rm ren}\ge 0\,,
\label{nposit}
\eeq
and
\beq
{\rm Re}\,\lambda \ge 0\, .
\label{lposit}
\eeq
The first condition just follows from $|n|^2\ge 0$, see (\ref{leq}).
The
second one becomes clear if we examine   the original bare action of
the model. From Eq.~(\ref{model}) we conclude
that the integral over $\lambda$ runs
along the imaginary axis (remember, we use the Euclidean formulation
of the theory.)
The saddle point solution for $\lambda$ can have (and will have) a
non-vanishing real
part. To ensure convergence of the path integral over
$n^{\ell}$'s
the real part of $\lambda$ at the saddle point should be
non-negative. It is important that $\sigma$ is an
independent integration variable, 
and the integral over $n^{\ell}$'s must be convergent at all values of 
$\sigma$, in particular at $\sigma=m_\ell$.

We will see shortly that the constraints
 (\ref{nposit}) and (\ref{lposit}) are important conditions which
single out physical phases existing in the given range of
the parameter $m$. In the subsequent sections we will study  solutions to our
master-set equations and show that
at large $m$ the theory is in the Higgs phase while at small $m$ it is
in the Coulomb/confining phase, the boundary being at $m=\Lambda$.

\subsection{The Higgs phase at large $N$}
\label{thehiggsp}

At  $m\gg \Lambda$ the solution to Eqs.~(\ref{leq}), (\ref{neq}) and
(\ref{seq})
has the following form:
\beqn
&& \langle \lambda \rangle =0\,,
\nonumber\\[2mm]
&&
\langle \sigma\rangle =m_0\, ,
\nonumber\\[2mm]
&&
\langle n \rangle = \sqrt{r_{\rm ren}}\, ,
\label{hvac}
\eeqn
where we use the gauge freedom of the original model to choose $n$
real, as was explained in Sect.~\ref{thehiggs}.
We see that the fields $\sigma$ and $n$ have non-vanishing VEV's and, as
a result, the $Z_N$ symmetry
is spontaneously broken. Our choice $n\equiv n_0$ was of course
arbitrary.
In fact,  we have
$N$ strictly degenerate vacua as shown in Eq.~(\ref{higgsvac}).
Equation~(\ref{seq}) must be used to fix the value of the constant
$c$,
\beq
c=\frac{1}{N}\,
\sum_{\ell=1}^{N-1}\left(1-\frac{m_{\ell}}{m_0}\right)
\ln\, {\frac{ |m_{\ell}-m_0|^2}{\Lambda^2}}.
\label{c}
\eeq
Substituting this value   in  the effective action (\ref{effaction}),
together with VEV's (\ref{hvac}), we get the vacuum energy in the
Higgs phase,
\beq
E_{\rm Higgs\,\, vac} =\frac{N}{2\pi}\, m^2\, .
\label{higgsenergy}
\eeq
The logarithmic term in the second line cancels the third line.

Now, let us have a closer look at the additional constraints
(\ref{nposit}) and (\ref{lposit}).
The latter condition  is trivially satisfied
while to examine the impact of the condition  (\ref{nposit}) we
substitute
$\sigma=m_0=m$ and  $\lambda=0$ in the expression (\ref{coupling})
for the renormalized coupling constant. Then we get
\beq
r_{\rm ren}=\frac1{4\pi}\, \sum_{\ell=1}^{N-1}
\ln\, {\frac{|m_{\ell}-m_0|^2}{\Lambda^2}}=
\frac1{\pi}\, \sum_{\ell=1}^{N/2}
\ln\, {\frac{2m \sin{\frac{\pi \ell}{N}}}{\Lambda}}
= \frac{N}{2\pi}\, \ln\, {\frac{m}{\Lambda}}\, ,
\label{hcoupl}
\eeq
where  the sum over $\ell$ is calculated in the large $N$ limit.
The constraint (\ref{nposit}) implies
\beq
m\ge\Lambda\,.
\label{higgsregion}
\eeq
The Higgs phase has a clear-cut meaning at large $m$. Hence, the
above result is compatible with intuition.
We will see momentarily that the lower bound of the allowed domain,
$m=\Lambda$, is the phase transition point.

\subsection{The Coulomb/confining phase at large $N$}
\label{theccp}

At small $m$ the appropriate solution of the master equations
(\ref{leq}), (\ref{neq})
and
(\ref{seq}) has the form
\beqn
&&
\sigma=0 \, ,
\nonumber\\[2mm]
&&
n=0\, ,
\nonumber\\[2mm]
&&
\lambda= \Lambda^2 -m^2\, .
\label{cvac}
\eeqn
The vacuum expectation value of the
$n$ field vanishes, as one would expect from the
$Z_N$ symmetric phase,  while Eq.~(\ref{leq}) is satisfied because
\beq
r_{\rm ren}=0
\label{ccoupl}
\eeq
in the vacuum (\ref{cvac}), cf. Eq.~(\ref{coupling}).
In fact Eq.~(\ref{cvac})
becomes an $m\neq 0$ generalization of Witten's saddle point condition
which was used to determine VEV of $\lambda$ in \cite{W79}.
Upon consulting with Eq.~(\ref{model})
we conclude that in our saddle point the mass of the $n^\ell$ quanta
is $\Lambda$, independent of the value of the mass deformation
parameter
$m$. Indeed, the mass squared $\to \lambda +|\sigma
-m_{\ell}|^2=\Lambda^2$.
Although this statement might seem counter-intuitive, it is correct.
We will comment on that in the end of this section.
Since  both $\sigma$ and $n$ do not condense in this
Coulomb/confining
 vacuum the
$Z_N$ symmetry is unbroken. The bare condition
(\ref{lambdaco}) is relaxed due to (\ref{ccoupl}). The solution
exhibits more degrees of
freedom than are present in the Lagrangian.

Let us turn now to constraints (\ref{nposit}) and (\ref{lposit}).
The first one is satisfied trivially while the second one implies
\beq
m\le\Lambda.
\label{coulombregion}
\eeq

We see that at $m\le\Lambda$ the theory is in the  Coulomb/confining
vacuum (\ref{cvac})
while at $m\ge\Lambda$ it is in the Higgs vacuum (\ref{hvac}). The
value
\beq
m_{*}=\Lambda
\label{trpoint}
\eeq
is the phase transition, or critical point.

Let us calculate the vacuum energy in  the Coulomb phase.
Substituting the vacuum values (\ref{cvac})  in the action
(\ref{effaction})
and using expression  (\ref{c}) for the value of the constant $c$
we arrive at the vacuum energy
\beq
E_{\rm Coulomb\,\, vac}
=\frac{N}{4\pi}\,\left \{\Lambda^2 +m^2+
m^2\log{\frac{m^2}{\Lambda^2}}\right \}\,,
\label{coulombenergy}
\eeq
where the sums over $\ell$ are calculated  in the
large-$N$ limit.

\begin{figure}
\epsfxsize=8cm
\centerline{\epsfbox{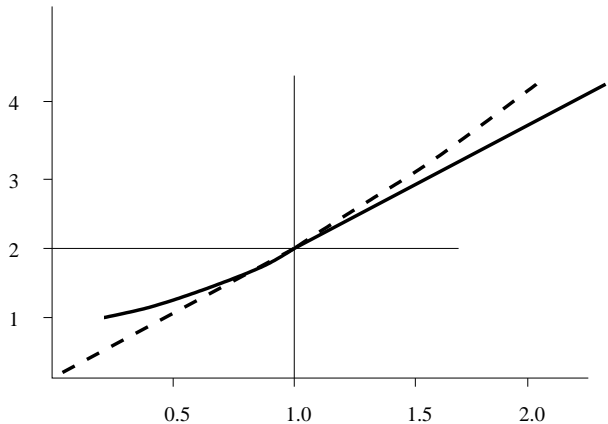}}
\caption{
Normalized vacuum energies $(4\pi E_{\rm vac}/N\,\Lambda^2)$ versus 
$m^2/\Lambda^2$. The solid line shows the actual vacuum energy,
while dashed lines correspond to a formal extrapolation of the Higgs and 
Coulomb/confinement
vacuum energies to unphysical values of $m$ below and above 
the phase transition point,
respectively.}
\label{fig:energies}
\end{figure}

We plot the vacuum energies (\ref{coulombenergy}) and (\ref{higgsenergy})
for the Coulomb/confining
and Higgs phases as a function of $m^2$ in
Fig.~\ref{fig:energies}. At the point of the phase transition
(\ref{trpoint}) energy densities of both phases coincide. Moreover,
their first derivatives
with respect to $m^2$ at this point coincide
too. The Higgs curve,
naively extrapolated below the phase transition, runs below the
 Coulomb curve which might lead one to conclude
 that the system
  always stays in the Higgs phase. However, the conditions
(\ref{nposit}) and
(\ref{lposit}) produce constrains (\ref{higgsregion}),
(\ref{coulombregion})
which tell us that at $m\le\Lambda$ the system is in the
Coulomb/confining
 phase while at
$m\ge\Lambda$ it is in the Higgs phase.

One can check our results for the vacuum energies in both phases performing
the calculations in a slightly different form, through
the trace of the energy-momentum tensor. The vacuum
energy can be obtained as
\beq
E_{\rm vac}=\frac{1}{2}\langle\theta_{\mu}^\mu\rangle
\eeq
where
\beq
\theta_{\mu}^\mu= \left[ M_{\rm uv}\partial_{M_{\rm uv}}
+\sum_{\ell=0}^{N-1}(m_l\partial_{m_\ell}
+ {m}_\ell^*\partial_{{m}_\ell^*})\right] {\cal L}(M_{\rm uv},m_\ell)\,,
\eeq
and ${\cal L}(M_{\rm uv},m_l)$ is the ultraviolet-regulated Lagrangian of the model.
Taking into account the classical contribution and the quantum anomaly, we precisely reproduce the vacuum energies
in the Coulomb/confining and Higgs phases quoted above.

To reiterate, at large $m$, at weak coupling, we have $N$ strictly
degenerate vacua;
the $Z_N$ symmetry is broken. At small $m$, at  strong coupling,
a mixing between these vacua takes over, and $N$ vacua split (see
Sect.~\ref{thecoulomb}). The order parameter which marks
these vacua is the  VEV  of the operator (\ref{topdens}) which
is non-zero for exited ``vacua'' with $k\ne 0$.
In the leading order in $N$
to which we are limited,  we do not see this vacuum splitting. Our
result exhibits
a single vacuum (\ref{cvac}). Moreover, we cannot say anything as  to
the nature of the phase transition. The answer can be found upon
inspection of a narrow strip
$|m^2-\Lambda^2|\sim O(1/N)$ (see Sect.~\ref{four}) which would
require tools going beyond those exploited here.

It is curious to note that the $\theta$ dependence of physical
quantities, albeit suppressed  at large $N$, 
is suppressed differently above and below the critical point.
If in the Higgs phase the suppression is expected to be exponential,
it is power-like in the Coulomb/confining phase.

Finally, we would like to comment on the independence of the
$n$-quanta mass on $m$ in the Coulomb/confining phase.
The crucial observation is that
in the absence of $n$ VEVs  the twisted mass term (\ref{Tthirteen})
is actually quartic in the $n$ fields, rather than quadratic.
Therefore, while it contributes to interactions of the $n$ fields,
its contribution to the $n$ mass cannot appear at order $N^0$.

\section{Can we describe critical behavior?}
\label{four}

The full solution of the phase transition problem requires
establishing
a conformal field theory which governs dynamics at the
critical point. It is no accident that so far nothing has been said
regarding this issue.
In this section we argue that the proper description of the critical
behavior
would require methods going beyond the $1/N$ expansion on which we
rely. Thus, this
question remains open.

To understand the nature of the phase transition one must identify
states that become massless at the critical point $m=\Lambda$.
Let us undertake this endeavor, approaching the critical point from
the Higgs side.

In the Higgs phase, at large $m$, the theory is weakly coupled,
and all excitations are massive. There are $N-1$ complex degrees of
freedom.
The lowest singularity in the correlation function
$\langle A_\alpha (x),\,\, A_\beta (0)\rangle$ is a two-particle
cut,
so that no stable field playing the role of a photon exist.
On the other hand, at $m=0$, there are $N$ quanta of the U(1)-charged
$n$ fields,
and a massless photon. It is natural to ask whether a light photon
emerges
at strong coupling as one approaches the critical point from above.

To address this issue we modify our effective action
(\ref{effaction}) including
the gauge field with a kinetic term induced at one loop as in
\cite{W79},
\beq
S_{\rm Higgs}=\int d^2 x\left \{ \frac{1}{4e_{\rm ren}^2}
F_{\alpha\beta}^2
+|\nabla_{\alpha}n|^2 +\frac{\tilde{e}_{\rm ren}^2}{2}\left( |n|^2-
r_{\rm ren}\right)^2
+ E_{\rm Higgs\,\,vac}\right\},
\label{heffaction}
\eeq
where $r_{\rm ren}$ is given in Eq.~(\ref{hcoupl}). The kinetic term
for the gauge field is induced through a loop of the $n^\ell$ quanta.
This loop converges in the infrared domain and, as we will see
shortly, is saturated by the lightest $n^\ell$ quanta,
i.e. $\ell \sim 1$.
In addition to this kinetic term, we also include a quartic term for
$n$ (we remind that $n\equiv n^0$) which comes from integration over
$\lambda$ in (\ref{effaction})
around the corresponding  saddle point at $\lambda=0$
(in the quadratic approximation). The corresponding
``coupling constants''  denoted by
  $e_{\rm ren}^2$ and $\tilde{e}_{\rm ren}^2$,
(in fact, they are momentum dependent; hence, the quotation marks)
are
\beq
e_{\rm ren}^2= \frac{12\pi}{\Sigma(p)}\,,\qquad
\tilde{e}_{\rm ren}^2=\frac{4\pi}{\Sigma(p)}\, ,
\label{couplings}
\eeq
where
\beq
\Sigma(p)=\sum_{\ell=1}^{N-1}\int\frac{ dk^2}{[k^2+|m_{\ell}-m_0|^2]
[(k-p)^2+|m_{\ell}-m_0|^2]}
\label{polop}
\eeq
and $p$ is the external momentum, $p_{\alpha} \leftrightarrow
i\p_{\alpha}$.

 The lightest states in the sum (\ref{polop})
 correspond to $\ell\sim 1$. Their
masses scale as
\beq
|m_{\ell}-m_0|^2 \sim \frac{m^2}{N^2}\quad\mbox{at}\quad \ell\sim
1\,.
\label{lightel}
\eeq
As long as the
 gauge field $A_{\alpha}$ and the scalar $n$ are much heavier, they
 cannot be treated as stable point-like bound states.

As we reduce $m$, the gauge field $A_{\alpha}$ and the scalar $n$
become lighter and
eventually may
cross the threshold and become genuinely stable bound states.
To evaluate their
masses let us take the low-energy limit in (\ref{polop}), assuming
that $p^2$ is
much less then the masses of lightest elementary states, see
Eq.~(\ref{lightel}).
Keeping only $\Sigma (0)$
we get from (\ref{heffaction}) the masses of the gauge field (the
massive gauge field
has one real degree of freedom) and the scalar $|n|$ (the phase of
$n$ is eaten
by the Higgs mechanism), respectively,
\beqn
&&\left.
m_{\gamma}^2=2\,r_{\rm ren}\,e^2_{\rm ren}\right|_{p^2=0}\,,\qquad
\left.
 m_n^2=2\,r_{\rm ren}\,
\tilde{e}^2_{\rm ren}\right|_{p^2=0}\,,
\label{massphn}
\\[3mm]
&&\Sigma(0)=0.17\,\frac{N^2}{2m^2}\,,
\label{sigma0}
\eeqn
where we calculated the sum over $\ell$ in (\ref{polop}) at $p=0$
numerically
in the limit of large $N$. The renormalized coupling constant $r_{\rm
ren}$
is given in Eq.~(\ref{hcoupl}).

As $m$ approaches $m_*=\Lambda$ from above,
the coupling constant $r_{\rm ren}$
tends to zero and, seemingly, so do the masses of the gauge field
$A_{\alpha}$ and scalar $n$,
\beq
m_{\gamma}^2 \sim m_n^2\sim \frac{\Lambda}{N}\,\delta m,
\label{bcmasses}
\eeq
where $\delta m=m-m_*$.
However, these conclusions would be  correct only if the masses
of these bound states were  much smaller than the masses of the
lightest
$n^\ell$ quanta. Comparing with
 (\ref{lightel}) we see that this would require
\beq
\frac{\delta m}{\Lambda}\ll \frac1N.
\label{region}
\eeq
In other words, the gauge field $A_{\alpha}$ and the scalar $n$
 could become light only
in a very close vicinity of the critical point. Unfortunately,
in the domain (\ref{region}) the expansion in
$1/N$, on which we heavily rely, explodes,
and we cannot trust our analysis.

%Moreover, if we look at  kink masses
%\beq
%\label{kinkmassren}
%M^{kink}_{\ell} \sim r_{ren}\,M_\ell
%\eeq
%near the point of the phase transition we see that they seems to go
%to zero
%too. The lightest masses ( at $\ell\sim 1$) scales as
%\beq
%(M^{kink}_{\ell\sim 1})^2\sim \delta m^2.
%\label{lightkinks}
%\eeq
%We see that at  the border of the region (\ref{region}) masses of
%lightest
%kinks become of the same order as masses of bound states
%(the gauge field $A_{\alpha}$ and scalar $n$ ). Again we cannot
%trust this
%analysis inside the region (\ref{region}) where $1/N$ corrections
%blow up.

Summarizing, in the narrow strip (\ref{region})
near the critical  point where light composite states could occur --
those which could become massless at criticality -- the $1/N$
expansion fails.
As a result, we cannot derive
 the conformal field theory which would describe our system at
criticality.

\section{Conclusions}
\label{conclu}

The mass deformed non-supersymmetric two-dimensional $CP(N-1)$ model,
with a special $Z_N$ preserving twisted mass term,
surfaced recently in connection with non-Abelian strings in
four-dimensional gauge theories \cite{GSY05}. This model turns out to
be  very interesting
on its own, as a theory with two distinct phases and a critical point
at strong coupling.
Using the large-$N$ expansion we confirmed the fact of the phase
transition in $m$,
determined the position of the critical point and calculated the
vacuum energies
in the Higgs and Coulomb/confining phases. The major unsolved problem
is determination of the conformal field theory governing dynamics
of the model at criticality.

The use of the large-$N$ expansion allowed us to bypass such
question as ``what particular aspect of the strong coupling dynamics
is responsible
for the change of regimes at $m=m_*=\Lambda$?" Although no definite
answer to this question can
be given at the moment, it is tempting to conjecture
that the phase transition is due to the fact that
at $m<m_*$ melted instantons of Fateev et al. \cite{fateev}
play a crucial role while at $m>m_*$
instantons are ``individualized" and suppressed.

Instantons are not suppressed at large $N$ at $m=0$
due to a large entropy factor.
The theory can be rewritten as a
massive fermion theory \cite{fateev} or, equivalently, as
the affine Toda theory at fixed coupling constant. Hence it is
natural to ask how our large-$N$
one-loop calculation captures nontrivial instanton effects.
A possible answer can be inferred from the supersymmetric version.
Supersymmetric theory can be treated in two different
ways \cite{hori}. Within the first approach one exploits a 
similar one-loop calculation, while within the second approach 
summation over nonperturbative
configurations yields a twisted superpotential of the affine Toda type. The
mirror symmetry of supersymmetric version is responsible for
equivalence of the vacuum structure in both approaches. In our
non-supersymmetric version
there is no evident notion of the mirror symmetry. However, one could
still hope that the relation between   instanton calculus and
the one-loop calculation in the linear gauged  
formulation of $CP(N-1)$  works in a similar manner.

The phase transition at some value of the twisted
mass we demonstrate in the present paper in $CP(N-1)$ seems to be a  
more general phenomennon taking place in a class of
asymptotically free non-super\-symmetric sigma models. In particular
such phase transition could be expected in the Grassmannian sigma models
as well as for many toric target manifolds. To an extent,
these phase transitions can be considered as  non-super\-symmetric
counterparts of the curves of marginal stability in supersymmetric
versions.

\section*{Acknowledgments}

We are  grateful to Adam Ritz for useful discussions.  
The work of A.G. was supported in part by grants CRDF RUP2-261-MO-04
and RFBR-04-011-00646.
A.G. would like to thank FTPI, University of Minnesota,
where a part of this work has been carried out, for the kind hospitality
and support. The work of M.S. work was supported in part by DOE 
grant DE-FG02-94ER408.  The work of A.Y. was  supported 
by  FTPI, University of Minnesota, and by Russian State Grant for Scientific School RSGSS-11242003.2.

\end{document}